\documentclass[12pt]{article}
\usepackage{graphicx}
\usepackage{amsmath}
\usepackage{hyperref}

\newcommand\pubdate{\today}

\def\Title#1{\begin{center} {\Large #1 } \end{center}}
\def\Author#1{\begin{center}{ \sc #1} \end{center}}
\def\Address#1{\begin{center}{ \it #1} \end{center}}

\newcommand\pubblock{\rightline{\begin{tabular}{l}  \\ 
         \pubdate  \end{tabular}}}
\newenvironment{Abstract}{\begin{quotation}  }{\end{quotation}}
\newenvironment{Presented}{\begin{quotation} \begin{center} 
             PRESENTED AT\end{center}\bigskip 
      \begin{center}\begin{large}}{\end{large}\end{center} \end{quotation}}

\begin{document}
\begin{titlepage}
 \pubblock
\vfill
\Title{Forward proton physics at LHC}
\vfill
\Author{Rafał Staszewski}
\Address{Henryk Niewodniczański 
Institute of Nuclear Physics\\
Polish Academy of Sciences \\
ul. Radzikowskiego 152 
31-342 Kraków, Poland}
\vfill
\begin{Abstract}
Diffractive phenomena constitute a large fraction of interactions occurring in $pp$ collisions at LHC. Because of the non-perturbative nature, their present understanding is still relatively poor and uncertain. One of the methods to study these processes is forward proton tagging. I will discuss the mechanism of diffractive processes, recent results, and potential implications. The proton tagging method can also be used for measurements of photon-induced processes, in particular, the photon–photon interactions. I will present the physics behind these processes, the experimental status and the lessons we can learn for the strong interactions and for the electroweak sector.
\end{Abstract}
\vfill
\begin{Presented}
DIS2023: XXX International Workshop on Deep-Inelastic Scattering and
Related Subjects, \\
Michigan State University, USA, 27-31 March 2023 \\
     \includegraphics[width=9cm]{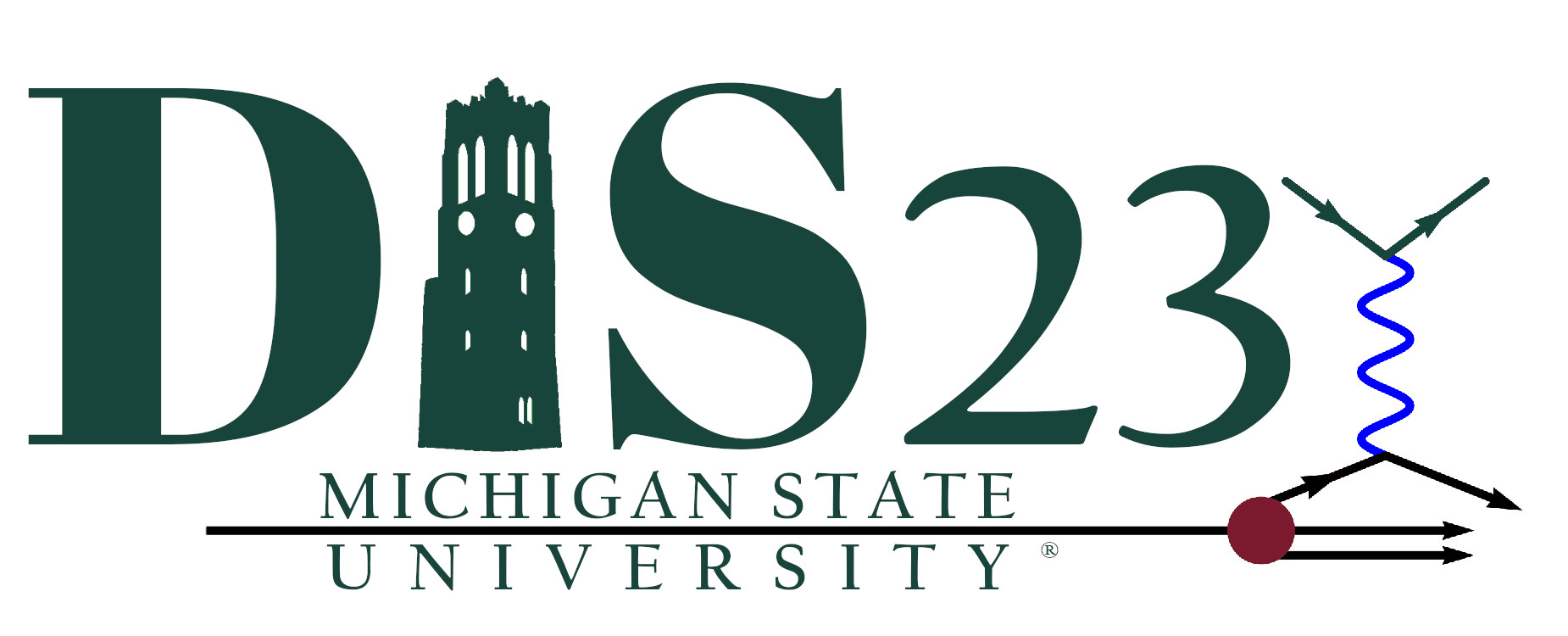}
\end{Presented}
\vfill
\end{titlepage}

\section{Introduction}

In most $pp$ interactions at high energies, the colliding protons are broken up by colour-octet exchanges occurring between their constituents. 
However, in some processes, these protons interact coherently and survive the interaction intact. 
At the LHC, such protons emerge from the interactions scattered at very small angles into the accelerator beam pipe. 
If their trajectory is at a sufficient distance from the beam orbit, it is possible to register them using detectors placed very close to the beam.
This is achieved using the technique of Roman pots.

\begin{figure}[hb]
\includegraphics[width=\textwidth]{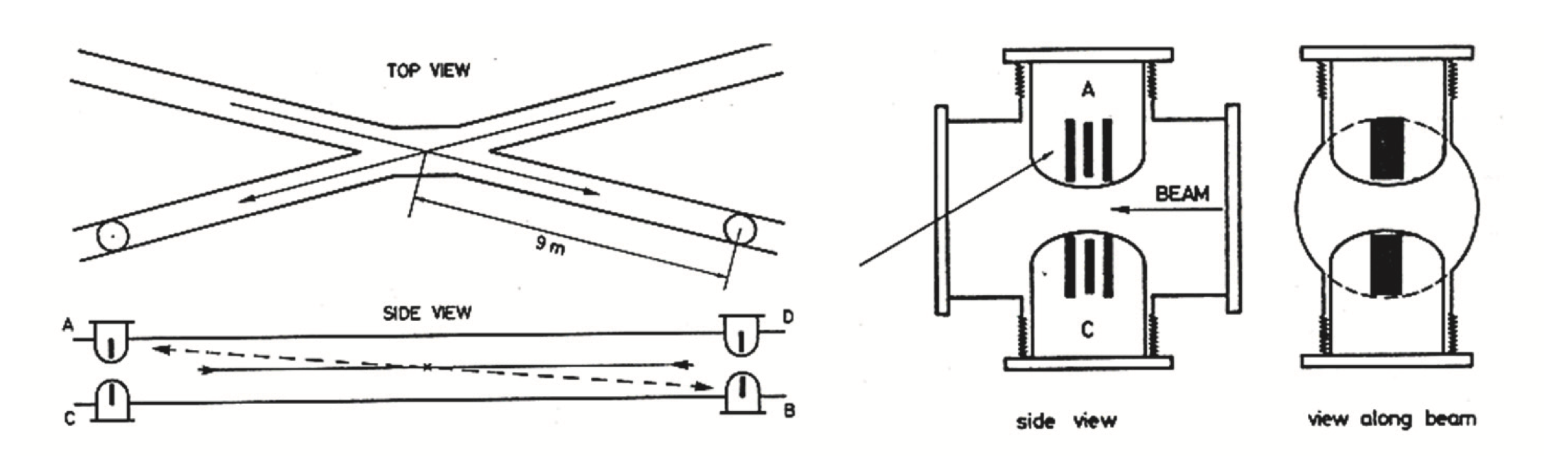}
\label{fig:rp}
\caption{Schematic illustration of the experimental set-up involving the Roman pots at ISR. Source: \cite{Fabjan:2017mno}}
\end{figure}

Roman pots take their name after the CERN-Rome group, which employed them for the first time at the ISR accelerator \cite{Fabjan:2017mno}. 
The experimental set-up consisted of stainless-steal vessels (pots) that could be mechanically inserted into the accelerator beam pipe for the data taking and retracted afterwards, see Fig. \ref{fig:rp}.
This ensures the accelerator safety in periods where the stability of the beam may not be optimal. An additional safety factor is provided by the secondary vacuum inside the Roman pots.

A sufficient distance between the beam protons and the scattered proton must be an effect of a difference in their momenta. 
For practical purposes, one can distinguish two factors that lead to the possibility of detecting the proton: a sufficiently large scattering angle or a sufficiently large momentum loss.
During typical LHC running, the beams colliding at the main interaction points are highly focused in order to provide high instantaneous luminosity.
Such strong focusing leads to large angular divergences of the beams.
This limits the minimal scattering angles that allow the detection of protons, making measurements of protons without a sufficient energy loss practically impossible. 
The solution to this problem is dedicated running of the accelerator.

\newcommand{\bstar}{$\beta^\ast$}
The setting of the accelerator magnets, the so-called machine optics, is often labelled by the \bstar{} value, which expresses the ratio of the beam transverse size to its angular divergence. 
Standard LHC operations are characterised by the \bstar{} value at the ATLAS and CMS interaction point of few tens of centimetres. 
In special high-\bstar{} runs, this value is much larger -- tens or even thousands of metres.

\section{Elastic scattering}

Using high-\bstar{} optics allows measurements of elastic scattering: $pp\to pp$.
This process, despite its kinematic simplicity, is of great importance because of its relation to other processes via the optical theorem. This theorem connects the total hadronic proton--proton cross-section $\sigma_\text{tot}$ with the elastic scattering amplitude $f_N(t)$:
\[
\sigma_\text{tot} = 4\pi \text{Im} f_N(t = 0),
\]
where $t$ is the four-momentum transfer squared.
For the case of scattering at very small angles, $t$ is up to a very good approximation equal to $-p_T^2$.

\begin{figure}[htb]
\includegraphics[width=.57\textwidth]{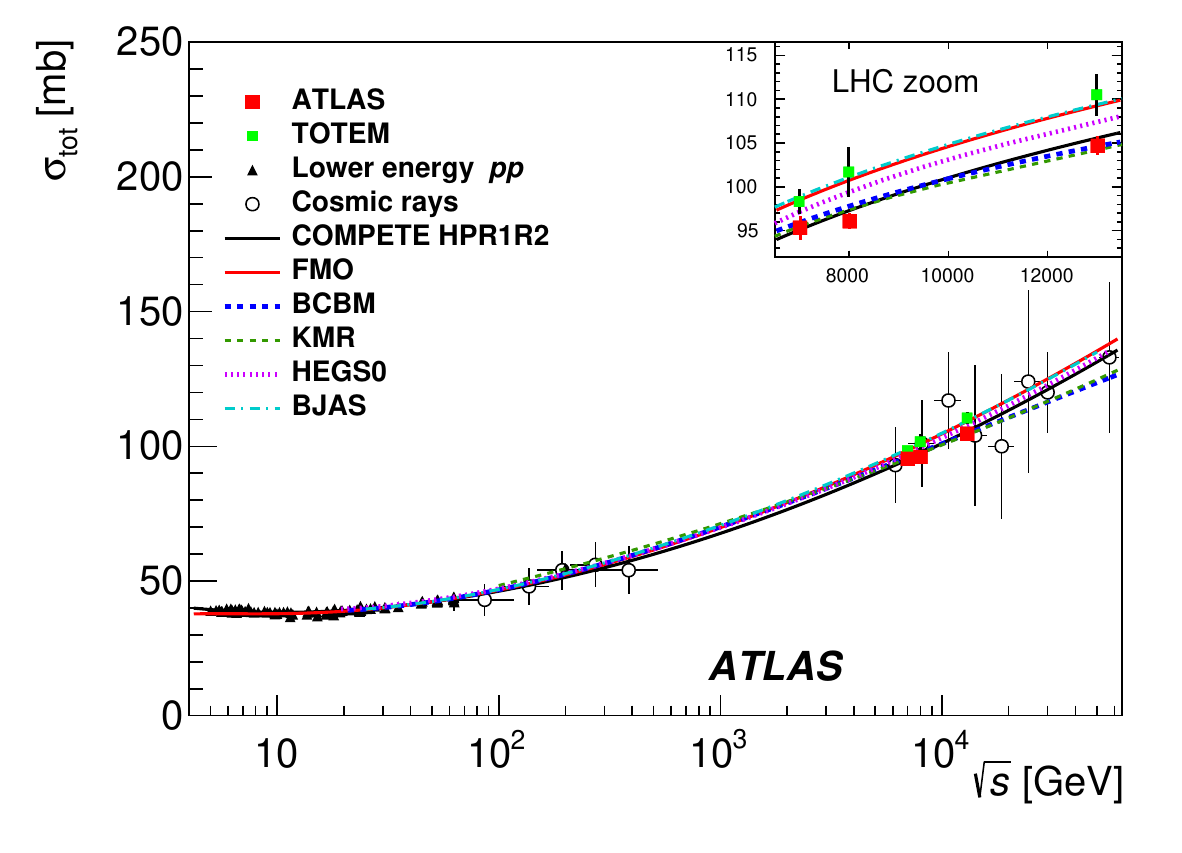}%
\includegraphics[width=.43\textwidth]{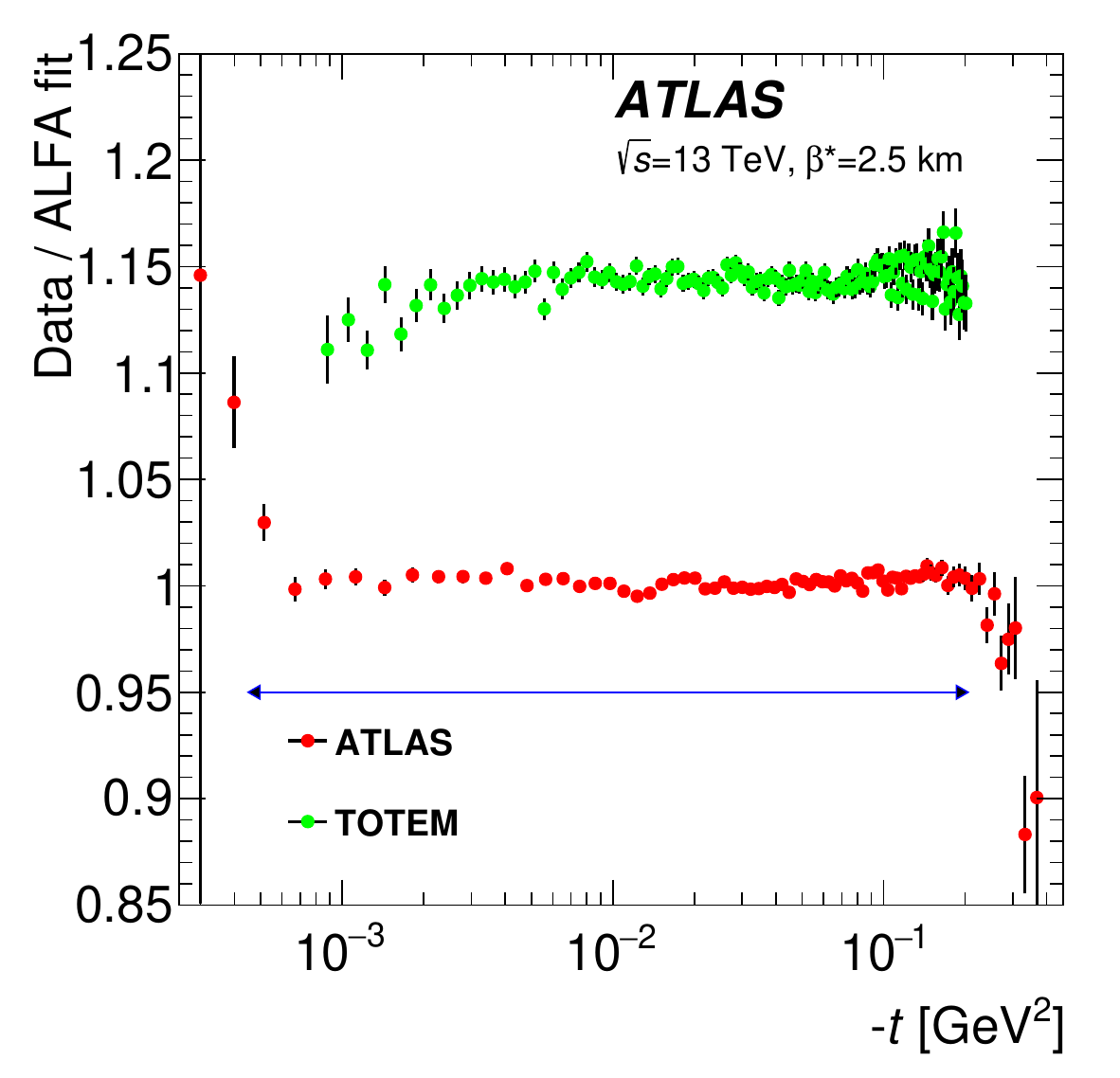}%
\caption{
Left: Comparison of differential elastic cross-section measured by ATLAS and TOTEM. 
Right: Summary plot of total proton--proton cross-section as a function of centre-of-mass energy.
Source: \cite{ATLAS:2022mgx}
\label{fig:sigtot}
}
\end{figure}

Fig \ref{fig:sigtot} (left) presents the measurements of $\sigma_\text{tot}$
from the ATLAS and TOTEM Collaborations at $\sqrt{s} = 13$~TeV \cite{ATLAS:2022mgx, TOTEM:2017sdy}.
It is important to mention that the measurements of the differential elastic cross-section from both experiments are in a good agreement concerning the shape but show a significant difference in the overall normalisation, see Fig. \ref{fig:sigtot} (right). 
This leads to a discrepancy in the $\sigma_\text{tot}$ values, which was also observed at 7 and 8 TeV. 
The exact reason for the discrepancy is unclear, but it is likely to be related to the analysis method -- ATLAS performed a~luminosity-dependent analysis, while TOTEM's analysis is luminosity-independent.

\begin{figure}[htb]
\includegraphics[width=.53\textwidth]{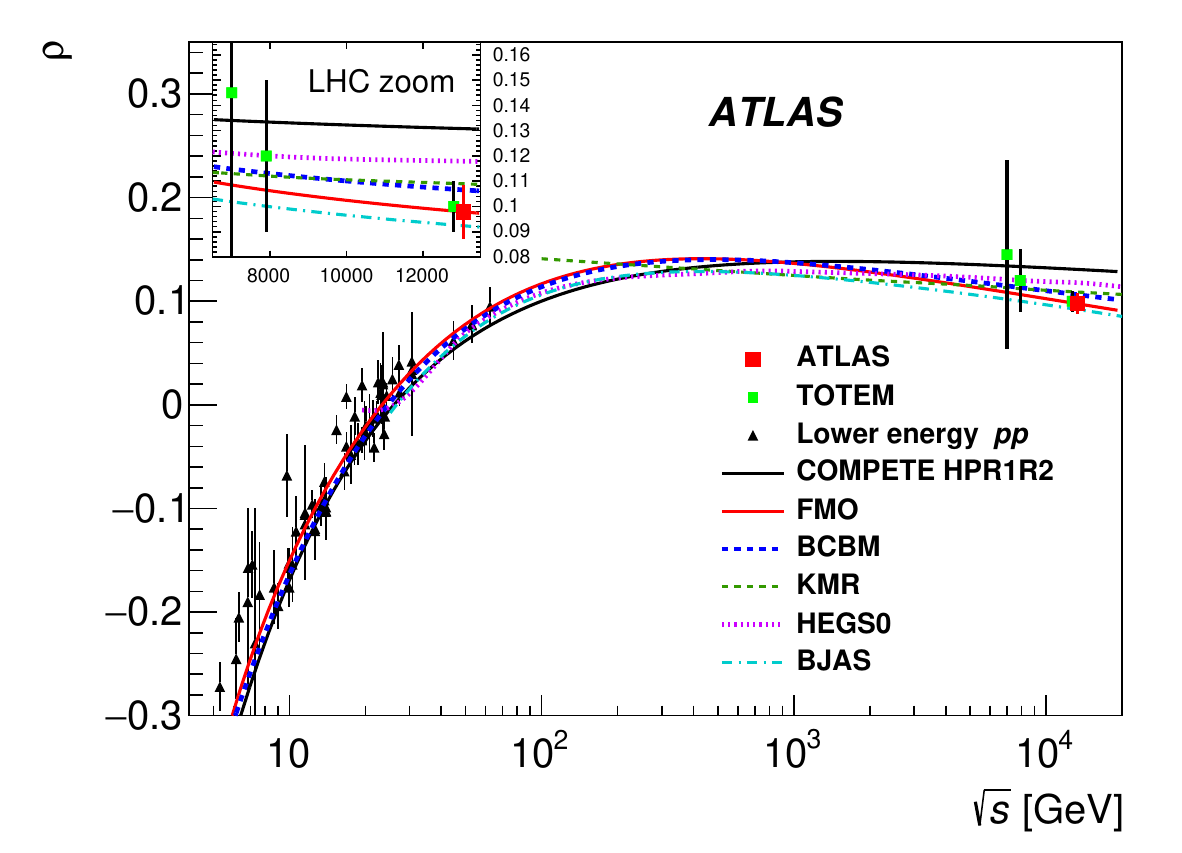}%
\includegraphics[width=.47\textwidth]{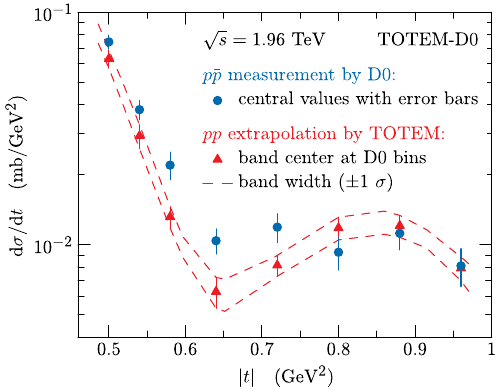}
\caption{
Left: Summary plot of $\rho$ as a function of the centre-of-mass energy;
from: \cite{ATLAS:2022mgx}.
Right: comparison of differential elastic scattering of $pp$ (extrapolated from TOTEM measurements) and $p\bar p$ (measured by D0) at $\sqrt{s} = 1.96$~TeV; from  \cite{TOTEM:2020zzr}.
\label{fig:rho}
}
\end{figure}

The interference between the nuclear and Coulomb mechanisms of elastic scattering provides sensitivity to the complex phase of the nuclear amplitude.
Fig. \ref{fig:rho} (left) presents the measurements of
\[
\rho = \left. \frac{\text{Re} f_N }{ \text{Im} f_N }\right|_{t=0}.
\]
The $\rho$ values measured by both experiments agree well.
Its value turned out to be significantly lower than the pre-LHC predictions \cite{COMPETE:2002jcr}.
It was proposed~\cite{TOTEM:2017sdy} that this difference is an effect of the odderon exchange.
This evidence was further strengthened by a combined analysis of the D0 and TOTEM data \cite{TOTEM:2020zzr}.
Using the TOTEM data, the differential elastic cross-section in the region of the diffractive dip measured at various LHC energies was extrapolated to the Tevatron energy. 
Comparison of these extrapolated data with the $p\bar p$ data measured by the D0 experiment showed a significant difference, see Fig. \ref{fig:rho} (right).
Based on both these arguments, the discovery of the odderon was claimed.

\section{Diffractive dissociation}
\begin{figure}[htbp]
\includegraphics[width=.58\textwidth]{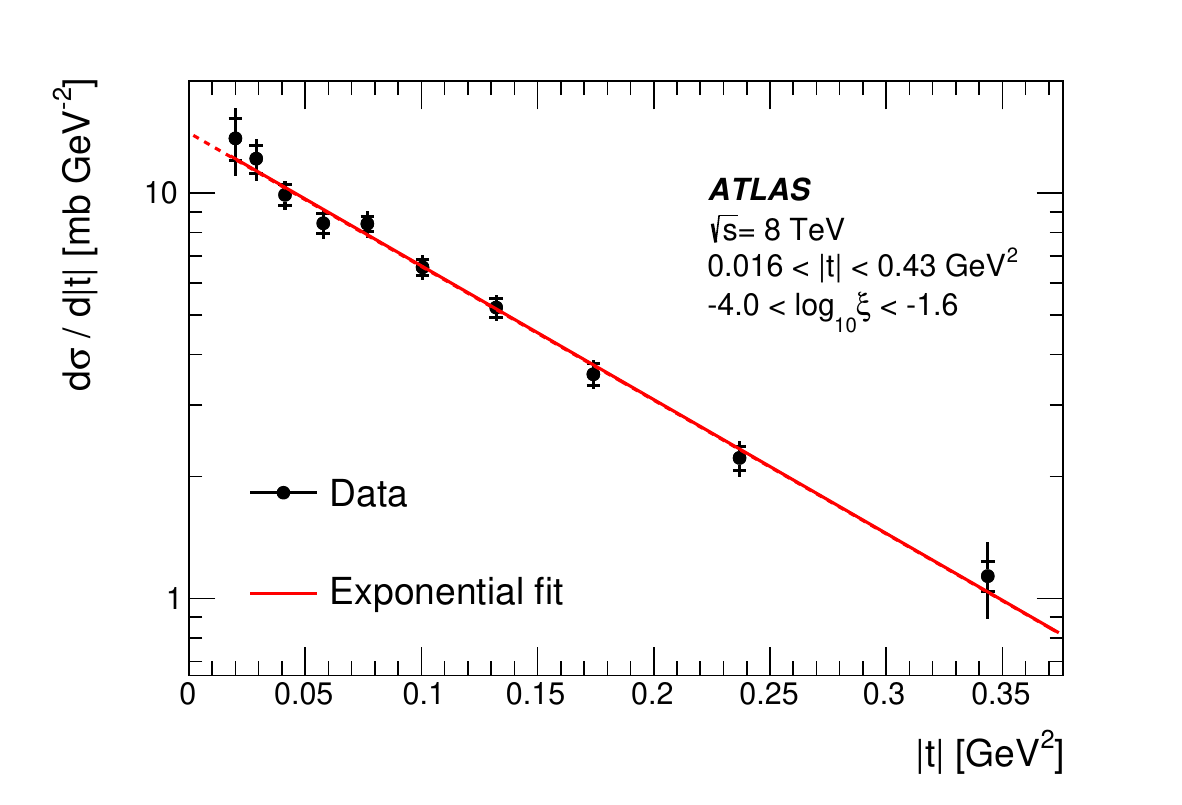}%
\includegraphics[width=.41\textwidth]{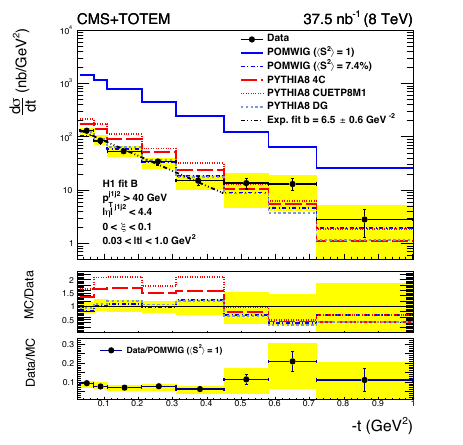}
\caption{
Left: 
the $t$ distribution in soft single-diffractive dissociation events as measured by ATLAS; 
from:
\cite{ATLAS:2019asg}.
Right: 
the $t$ distribution in hard single-diffractive dissociation events as measured by CMS and TOTEM; 
from:
\cite{CMS:2020dns}.
\label{fig:sd}
}
\end{figure}

High-\bstar{} runs can be used for studies of diffractive-dissociation processes, $pp\to pX$
In order to reduce background coming from beam halo, detection both the scattered proton as well as a part of the dissociated system $X$ are usually required.
The particular process to be measured depends on the requirements imposed on the $X$ state. 
Fig. \ref{fig:sd} (left) shows the result of the ATLAS measurement of soft single-diffraction.
Fig.~\ref{fig:sd} (right) presents the results of the CMS-TOTEM measurement of single-diffraction events containing high-$p_T$ jets.

\section{Central exclusive production}
\begin{figure}[htbp]%
\includegraphics[width=.397\textwidth]{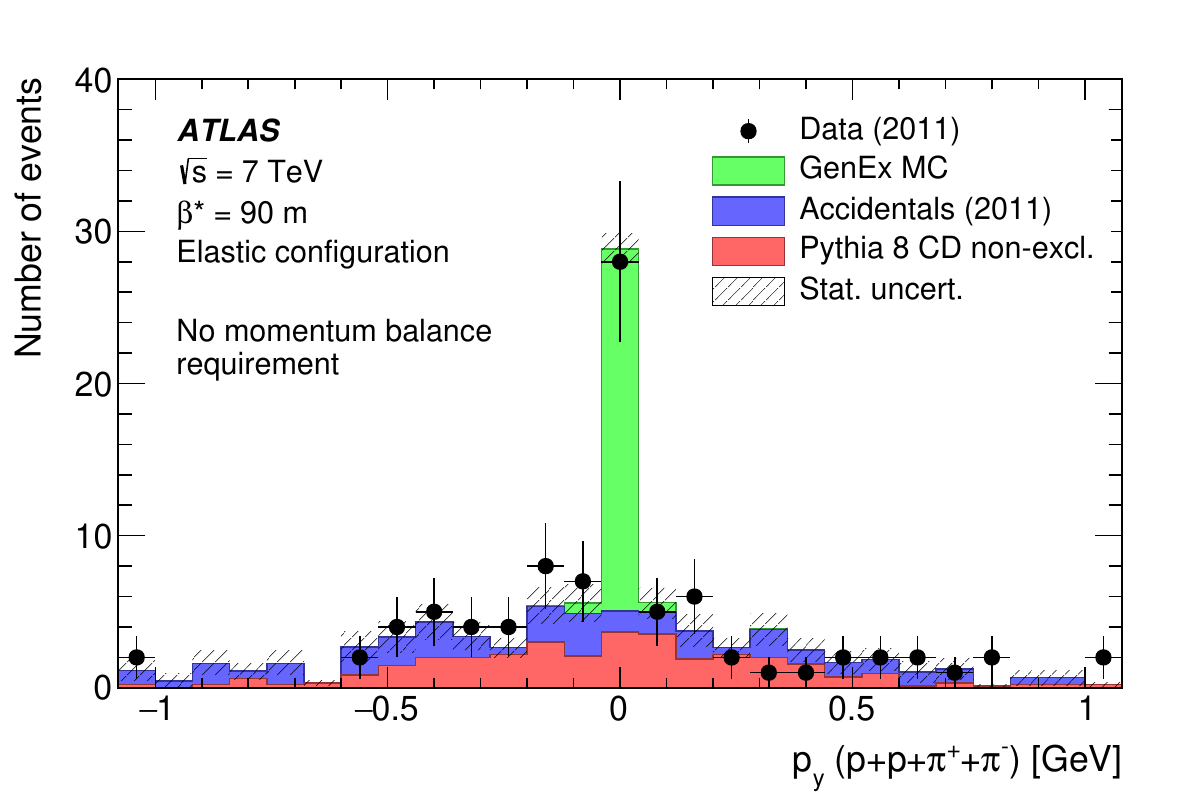}%
\includegraphics[width=.347\textwidth]{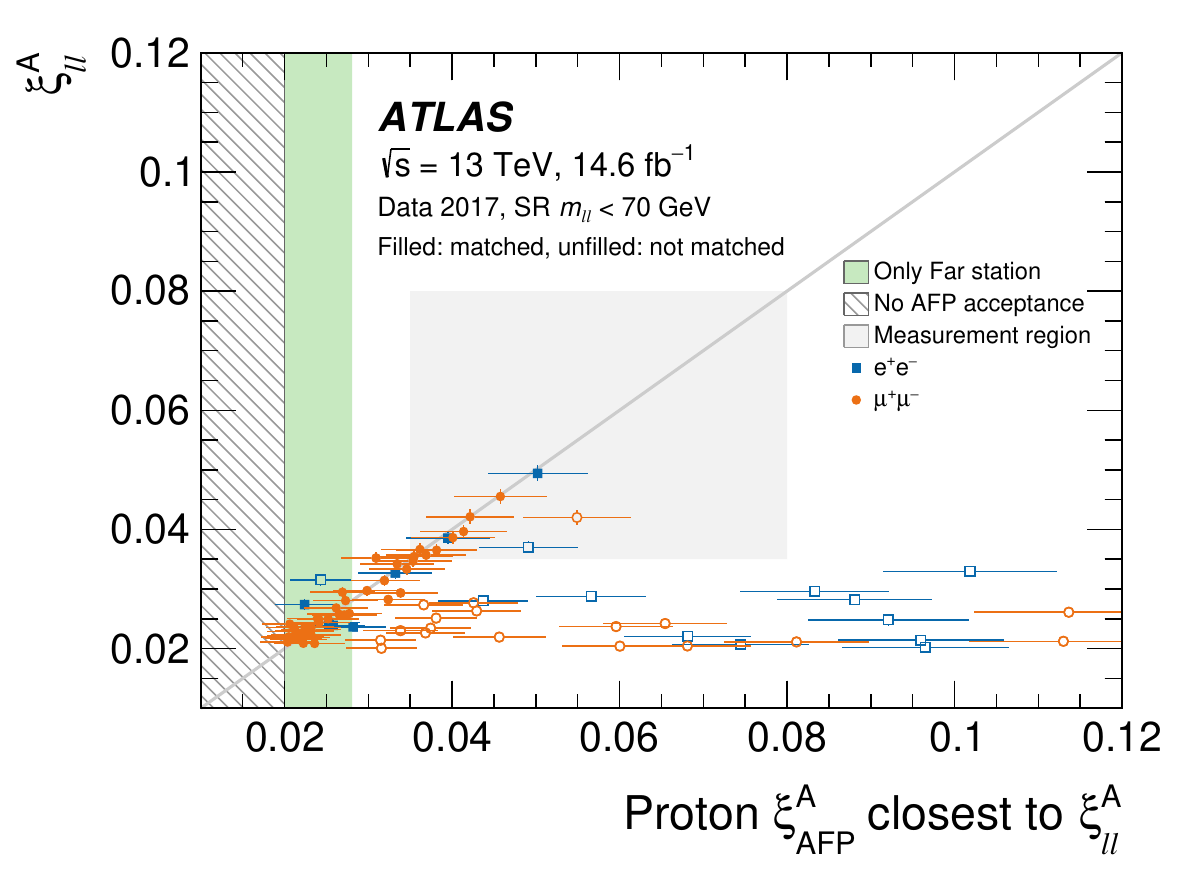}%
\includegraphics[width=.257\textwidth]{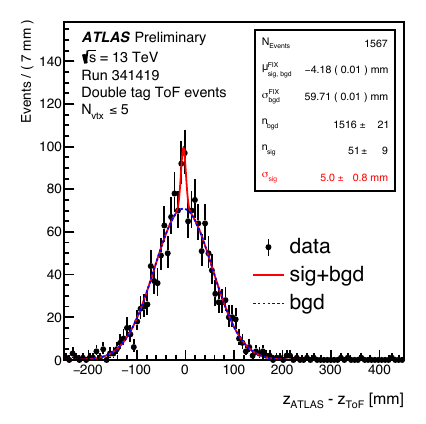}%
\caption{
Methods of rejecting non-exclusive background.
Left: based on the transverse momentum matching;
source:
\cite{ATLAS:2022uef}.
Centre: based on the longitudinal momentum matching; source:
\cite{ATLAS:2020mve}
Right: based on the vertex matching; 
source:
\cite{ToF}
}
\label{fig:exclusivity}
\end{figure}

Another interesting process to study is the central exclusive production, $pp\to pXp$, 
where both scattered protons and the centrally produced system $X$ can be detected. 
The first such measurement at LHC was reported by the ATLAS Collaboration and concerned 
the production of charged pion pairs: $pp\to p \pi^+\pi^- p$ \cite{ATLAS:2022uef}.
The exclusivity is ensured by vetoing additional activity in the forward direction and ensuring that the transverse momenta of all produced particles add up to zero, see Fig. \ref{fig:exclusivity} (left). 
With the limited statistics available, the main result is the fiducial cross-section. 
At the present moment, theoretical predictions for the fiducial volume of the measurement are not available. 
It will be interesting to see if the theory is able to describe the published data.

\section{Two-photon processes}

As anticipated earlier, processes with forward protons can be measured also during LHC standard running. 
However, for such low-\bstar{} conditions one is in practice limited to detecting those protons that separated from the beam as an effect of losing energy in the interaction and thus having their trajectory bent more in the LHC magnets (particularly in the pair of dipole magnets placed around 100 m from the interaction point, i.e. those which are used for separating the incoming and outgoing beams). 

Taking data in runs with high instantaneous luminosity allows studying rare reactions, in particular the two-photon processes, in which both protons emit quasi-real photons that interact with each other.
Those processes can be exploited for probing the electro-weak sector and searching for effects of new physics.
Such processes are also interesting from the point of view of the strong interactions because of the rescattering corrections that can spoil the signature of the event.

One of the main experimental problems is the combinatorial background.
It is related to high pile-up, \textit{i.e.} a large number of $pp$ interactions occurring simultaneously in the same event, and is a consequence of high instantaneous luminosity.
A two-photon event, for example $pp\to p\gamma\gamma p\to p l \bar l p$, can be mimicked by an event where the lepton pair is produced in a Drell-Yan process, and the protons are produced in two independent soft single-diffractive $pp$ collisions.

Forward proton detectors provide two ways to reject combinatorial background events.
In cases where full four-momentum of the centrally-produced state can be measured, it is possible to check its compatibility with the momentum of the forward proton. For example, in the $pp\to pl\bar l p$ process, the measurement of the $l\bar l$ system and assuming the two-photon production mechanism, one can obtain the expected proton longitudinal momentum.
It can then be compared to the value measured directly.
For genuine two-photon processes both values should be very close, while for combinatorial background they should be uncorrelated.
This is illustrated in Fig.~\ref{fig:exclusivity}~(centre).

The other method of background rejection requires a precise measurement of the arrival time of the forward protons.
Using the timing information from both protons, it is possible to calculate the longitudinal coordinate of the interaction position. 
This can in turn be compared to the measurement using the central tracking detectors, see Fig.  \ref{fig:exclusivity} (right).
This method has not yet been employed for any physics analysis, but it was shown to work in performance studies \cite{ToF}.

\begin{figure}[t]
\includegraphics[width=.5\textwidth]{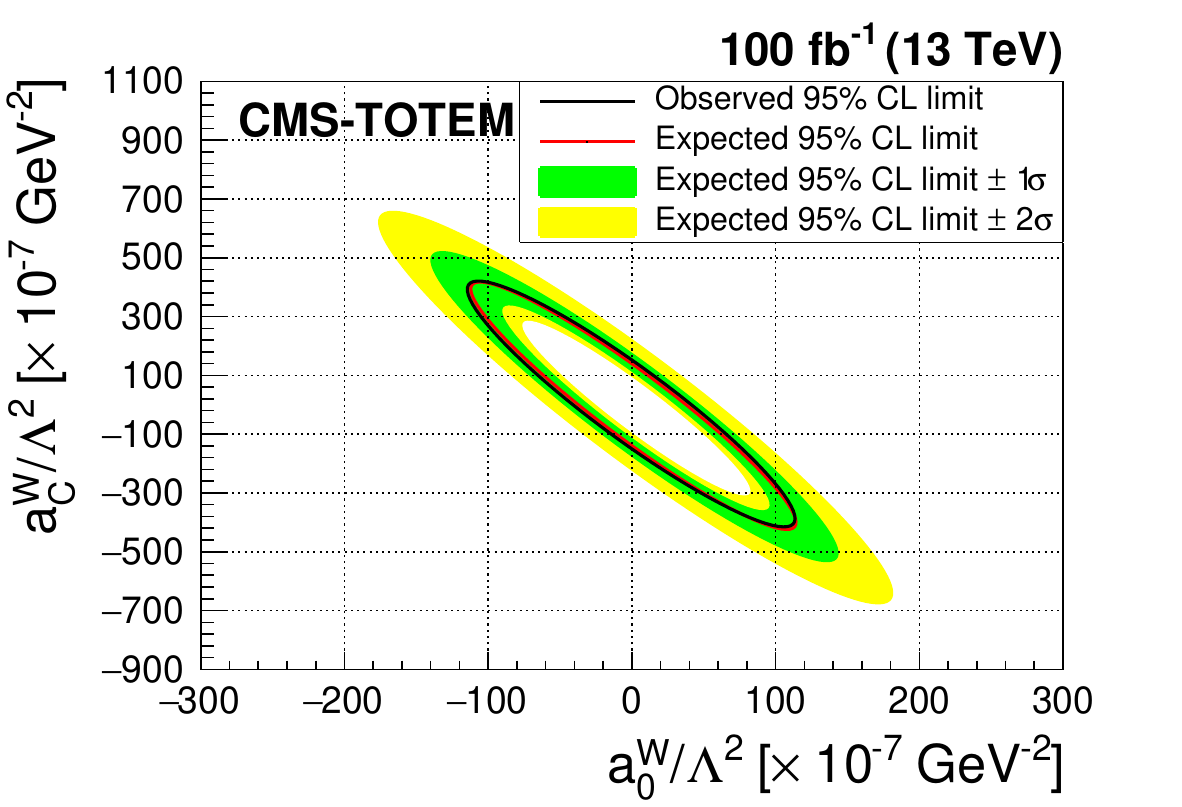}%
\includegraphics[width=.5\textwidth]{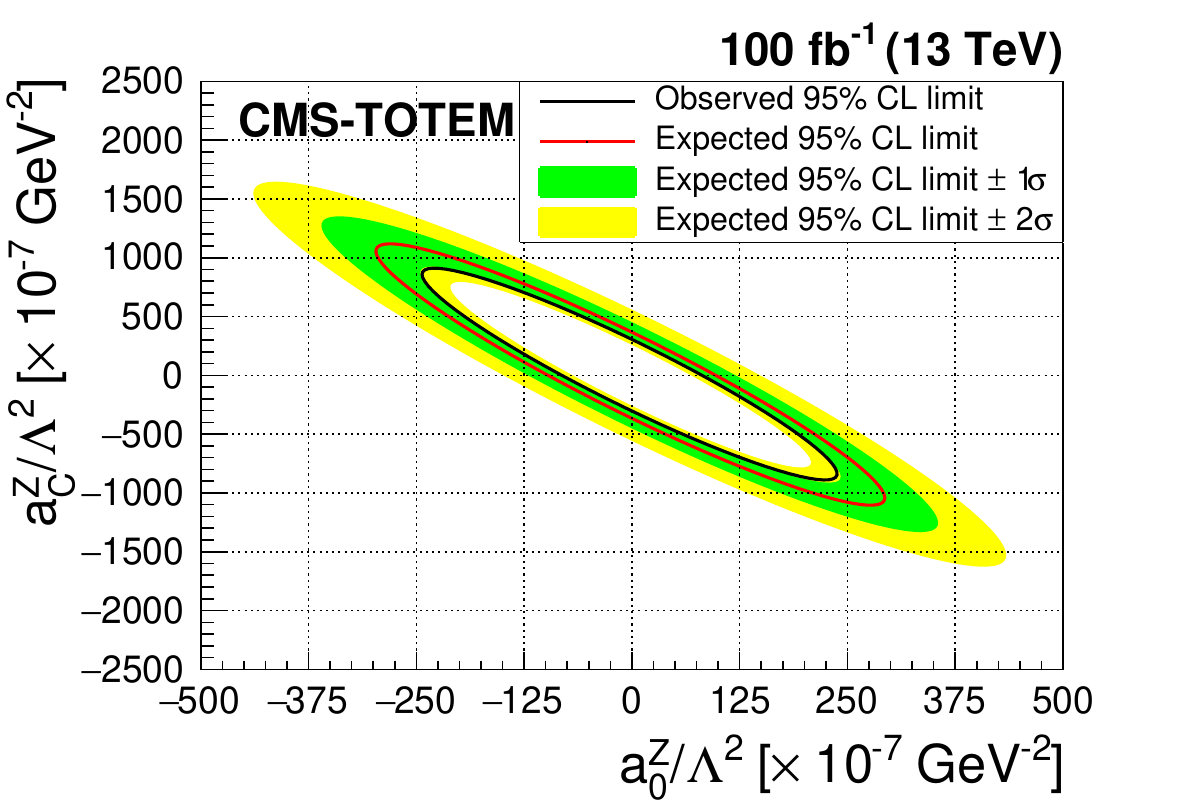}%
\caption{
Limits on the anomalous quartic gauge couplings obtained by CMS and TOTEM. Source:
\cite{CMS:2022dmc}
}
\label{fig:aqgc}
\end{figure}

\begin{figure}[t]
\includegraphics[width=.5\textwidth]{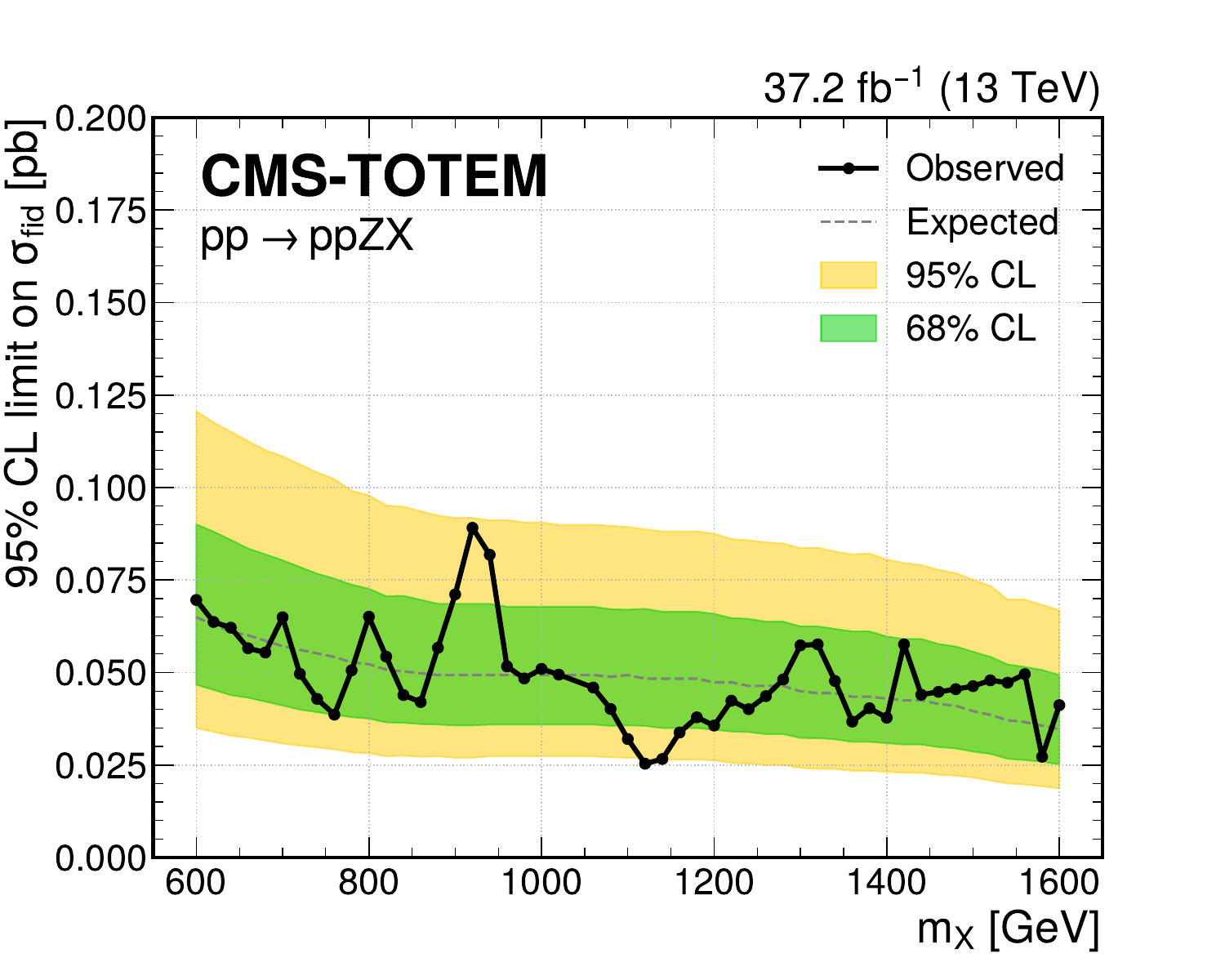}%
\includegraphics[width=.5\textwidth]{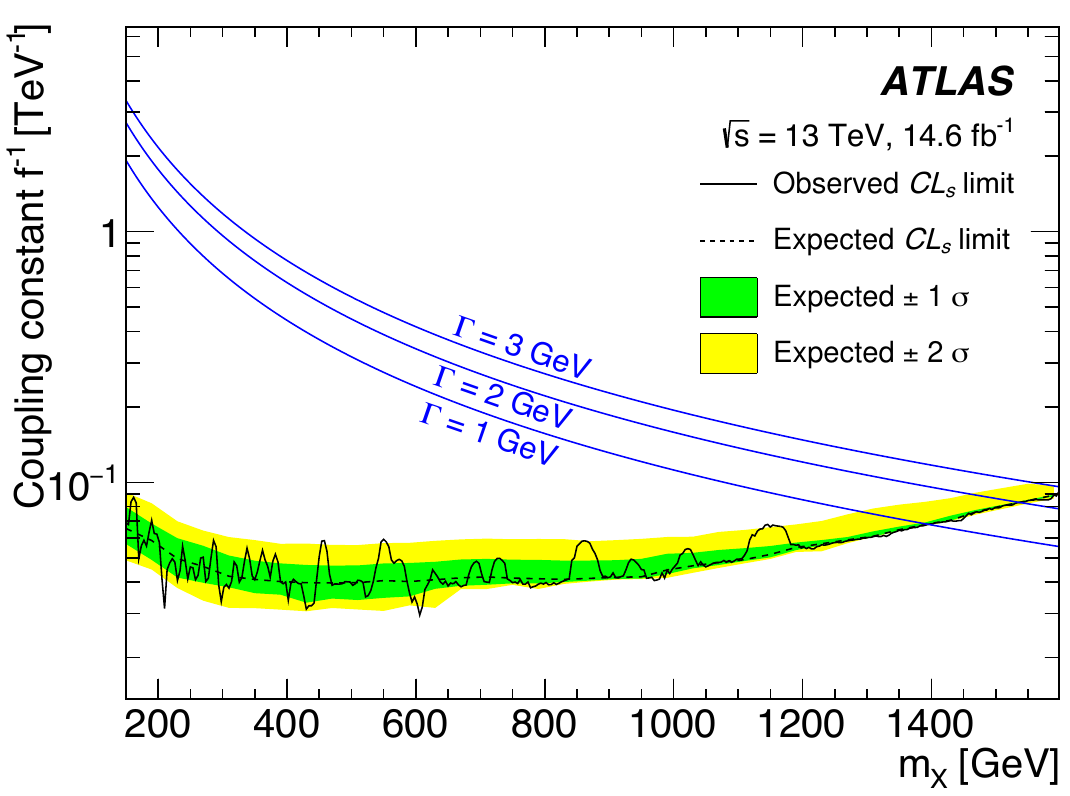}%
\caption{
Right: ATLAS limits for coupling of axion-like particles produced in the $s$-channel of $\gamma\gamma\to\gamma\gamma$;
source:
\cite{ATLAS:2023zfc}
Left: CMS-TOTEM cross-section limits for the production of an invisible particle in a two-photon process together with the $Z$ boson; source:
\cite{CMS:2023roj}.
}
\label{fig:ZXalp}
\end{figure}

The kinematic matching was used for several searches for BSM physics.
It allowed setting limits on the anomalous quartic couplings in the $\gamma\gamma\to WW$ and $\gamma\gamma\to ZZ$ processes \cite{CMS:2022dmc}, see Fig. \ref{fig:aqgc}.
It was also used to search for axion-like particles exchanged in the $s$-channel of the $\gamma\gamma\to\gamma\gamma$ process \cite{TOTEM:2021zxa,ATLAS:2023zfc}, see Fig. \ref{fig:ZXalp} (left).

Since knowing the kinematics of both forward protons constrains the kinematics of the centrally produced state, the forward proton detectors provide new ways of searching for invisible particles.
Recently, CMS and TOTEM showed a study of $\gamma\gamma\to ZX$, where $X$ is an invisible state.
By measuring the $Z$ and both protons it is possible to reconstruct the mass of $X$.
The resulting cross-section limits are shown in Fig. \ref{fig:ZXalp} (right).

\section{Summary and conclusions}

Forward proton tagging is an experimental technique that enriches the physics programme at the LHC. 
It allows measurements that would not be possible in other ways, in particular, in various studies of elastic scattering. 
For other, diffractive, and photon-induced processes, it can be used to reject background, add new observables to the analysis, or ensure exclusivity. 
Several measurements have already been performed, but many more are yet to be done, using data already collected as well as those to be taken in the future.

\section*{Acknowledgements}
This work is supported in part by Polish National Science Centre NCN SONATA BIS grant number 2021/42/E/ST2/00350.

\end{document}